# Fractal-based Correlation Analysis for Resting State Functional Connectivity of the Rat Brain in Functional MRI

*Wonsang You[1], Jörg Stadler[1]*
[1]The Special Lab Non-Invasive Brain Imaging, Leibniz Institute for Neurobiology, Germany

## Introduction

The most studies on functional connectivity have been done by analyzing the brain's hemodynamic response to a stimulation. On the other hand, the low-frequency spontaneous fluctuations in the blood oxygen level dependent (BOLD) signals of functional MRI have been observed in the resting state. However, the BOLD signals in resting state are significantly corrupted by huge noises arising from cardiac pulsation, respiration, subject motion, scanner, and so forth. Especially, the noise compounds are stronger in the rat brain than in the human brain. To overcome such an artifact, we assumed that fractal behavior in BOLD signals reflects low frequency neural activity, and applied the theorem such that the wavelet correlation spectrum between long memory processes is scale-invariant over low frequency scales. Here, we report an experiment that shows special correlation patterns not only in correlation of scaling coefficients in very low-frequency band (less than 0.0078Hz) but also in asymptotic wavelet correlation. In addition, we show the distribution of the Hurst exponents in the rat brain.

## Methods

The asymptotic wavelet correlation has been usually estimated with the bivariate least mean squares estimator of the Hurst exponent. However, when a time series has more than one pair for correlation analysis, it can also have different values of its Hurst exponent. To avoid this conflict, we exploited an univariate estimator to get the Hurst exponent individually at each region, and afterward compute the asymptotic wavelet correlation as illustrated at Fig. 1.

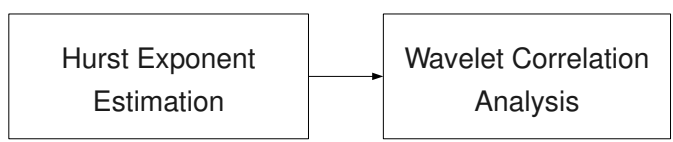

**Figure 1. The asymptotic wavelet correlation analysis.**

As illustrated at Fig. 2(a), the periodogram-based estimator of the Hurst exponent was applied since it is especially appropriate to analyze time series which have significant long memory property near zero frequency.

The asymptotic wavelet correlation between time series $X_1$ and $X_2$ can be computed by the least mean squares estimator that finds out the best scale range which satisfies the scale-invariance of the wavelet correlation. Let us assume that we have long memory parameters which satisfy the following linearity in the scale range

$$\log_2[\gamma_{X_1}(j)] = 2d_1 j + c_1, \quad \log_2[\gamma_{X_2}(j)] = 2d_2 j + c_2$$
$$\log_2[\gamma_{X_1,X_2}(j)] = (d_1+d_2)j + c_{12}.$$

where wavelet variances and covariance $\gamma_{X_1}, \gamma_{X_2}$, and $\gamma_{X_1,X_2}$ are given respectively in each scale. Then, the asymptotic wavelet correlation can be simply calculated by the following equation.

$$\rho_{\lim J\to\infty} = 2^{c_{12}-(c_1+c_2)/2}$$

Also, we computed the correlation of scaling coefficients which reflects transient change in very low-frequency range.

## Results

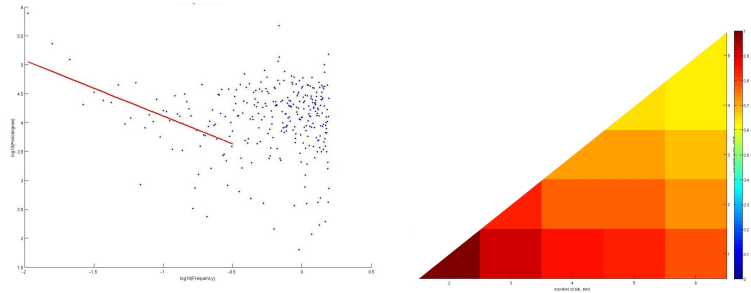

**Figure 2.** (a) The periodogram-based estimation of the Hurst exponent, and (b) the least mean square estimation of the covariance parameter.

Fig. 3 shows that the Hurst exponent is higher in subcortical regions –especially hippocampus– than in cortical regions.

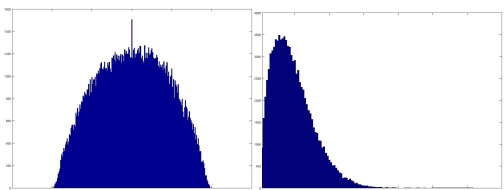

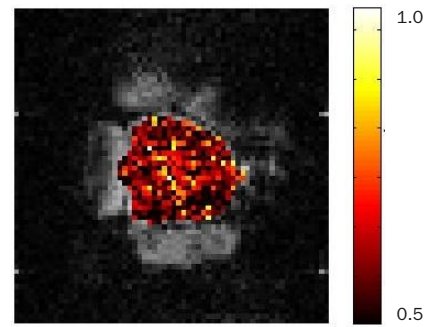

**Figure 4.** The histograms of (a) correlations of scaling coefficients and (b) asymptotic wavelet correlation.

**Figure 3.** The distribution of Hurst exponents

In Fig. 4, correlation coefficients of scaling coefficients are symmetrically but not normally distributed with short tails while the histogram of asymptotic wavelet correlations resembles a left-skewed gamma distribution.

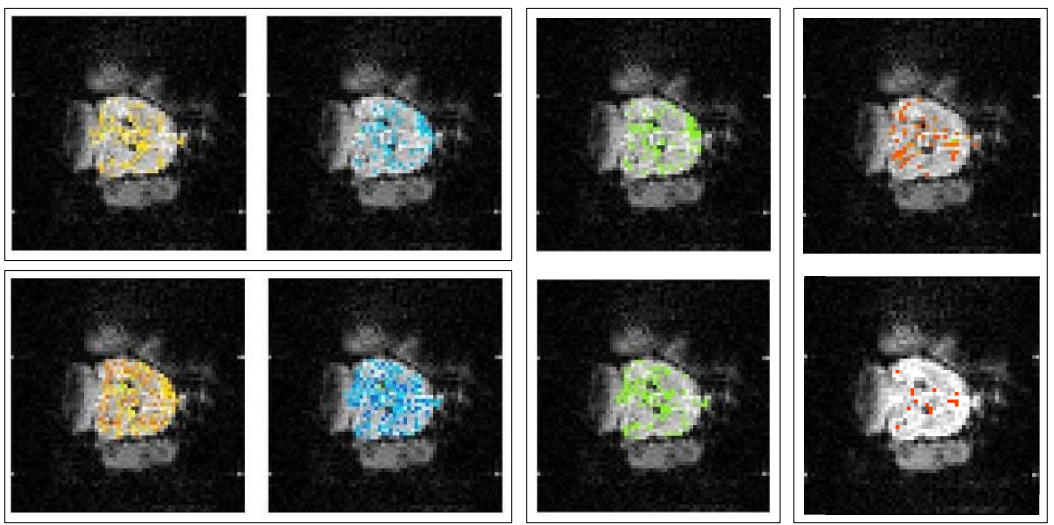

**Figure 5.** (a) (top-left) Y-type pattern in positive correlation and its corresponding O-type pattern in negative correlation, (b) (bottom-left) O-type pattern in positive correlation and its corresponding Y-type pattern in negative correlation. (c) (middle) The distributions of voxels with positive Y-type pattern and positive O-type pattern. (d) (right) Y-type pattern in the asymptotic wavelet correlation and the distribution of such voxels.

We found two special groups of voxels which have Y-type or O-type patterns in correlation of scaling coefficients. The Y-type in positive correlation is always followed by the O-type in negative correlation, and vice versa as shown at Fig. 5. Moreover, the voxels with the positive Y-type pattern are distributed like the O-type, and vice versa as shwon at Fig. 5(c). Likewise, the Y-type pattern had been observed even in the asymptotic wavelet correlation, as shown at Fig. 5(d).

## Discussion

The special patterns, both in correlation of scaling coefficients and the asymptotic wavelet correlation, need to be validated by other quantitative and neuroscientific ways. To avoid noise effect, modeling system and physiological noises will be instrumental to elaborately extract the endogenous signals by estimating noise parameters. Our future works include developing a robust method to distinguish the physiological noise with scale-invariant property from the endogenous signals.